
\input phyzzx
\nonstopmode
\sequentialequations
\twelvepoint
\nopubblock
\tolerance=5000
\overfullrule=0pt

\REF\ab{Y. Aharonov and D. Bohm, {\it Phys. Rev.} {\bf 115},
485 (1959).}

\REF\amrw{M. Alford, J. March-Russell and F. Wilczek,
{\it Nucl. Phys.} {\bf B328}, 140 (1989).}

\REF\promise{J. March-Russell, J. Preskill, and F. Wilczek,
in preparation.}

\REF\berry{M. Berry, {\it Proc. R. Soc. Lond.} {\bf A392},
45 (1984).}

\REF\tomita{A. Tomita and R. Chiao, {\it Phys. Rev. Lett.}
{\bf 57}, 937 (1986).}

\REF\berryopt{M. Berry, {\it Nature} {\bf 326}, 277 (1987).}

\REF\schwarz{A. Schwarz, {\it Nucl. Phys.} {\bf B208}, 141
(1982).}

\REF\cheshire{M. Alford, {\it et al.},
{\it Phys. Rev. Lett.} {\bf 64},
1632
(1990); and {\it Nucl. Phys.} {\bf B349}, 414 (1991).}

\REF\prekra{J. Preskill and L. Krauss, {\it Nucl. Phys.}
{\bf B341}, 50 (1990).}

\REF\ww{F. Wilczek and Y.-S. Wu, {\it Phys. Rev. Lett.}
{\bf 65}, 13 (1990).}

\REF\bucher{M. Bucher, {\it Nucl. Phys.} {\bf B350}, 163
(1991).}

\REF\kman{A possible analogue of the Aharonov-Bohm effect in
the context of helium 3 has been suggested by M. Khazan,
{\it JETP Lett.} {\bf 41}, 486 (1985).
He does not discuss the momentum dependence of his effect,
nor
does he display an induced potential.  It is unclear to us
whether his effect falls into the framework discussed here.}

\REF\mermin{N. Mermin, {\it Rev. Mod. Phys.} {\bf 51}, 591
(1979).}

\REF\volovik{G. Volovik and V. Mineev, {\it JETP} {\bf 45},
1186 (1977).}

\REF\giddings{S. Giddings, in {\it Quantum Coherence},
ed. J. Anandan (World Scientific, 1989).
The possibility of deriving black hole
hair from Aharonov-Bohm
type effects was first discussed in M. Bowick et al,
{\it Phys. Rev. Lett.} {\bf 61}, 2823 (1988).  However,
more recent work indicates that
the only measurable forms of quantum hair are related to
local gauge symmetries.  See S. Coleman, J. Preskill, and
F. Wilczek, {\it Mod. Phys. Lett.} {\bf A6}, 1631 (1991),
and in preparation.}

\def\Z{{Z}}
\let\al=\alpha
\let\be=\beta

\let\de=\delta

\let\del=\nabla
\let\si=\sigma

\let\th=\theta

\let\om=\omega

\let\p=\partial
\let\<=\langle
\let\>=\rangle

\def\comment#1{ \hbox{Comment suppressed here.} }

\def\rhp{\rho_1}
\def\rhm{\rho_2}
\def\mup{\mu_1}
\def\mum{\mu_2}

\def\km{k_2}

\line{\hfill PUPT-91-1297}
\line{\hfill IASSNS-HEP-91/92}
\line{\hfill CALT-68-1763}
\line{\hfill December 1991}
\titlepage
\title{Internal Frame Dragging and a Global Analogue
of the Aharonov-Bohm Effect}
\vskip.2cm
\author{John March-Russell\foot{Research supported by NSF
grant
NSF-PHY-90-21984. e-mail: jmr@puhep1.princeton.edu,
jmr@iassns.bitnet}}
\vskip.2cm
\centerline{{\it Joseph Henry Laboratories}}
\centerline{{\it Princeton University}}
\centerline{{\it Princeton, N.J. 08544}}
\vskip.2cm
\author{John Preskill\foot{Research supported in part by DOE
grant
DE-AC03-81-ER40050}}
\vskip.2cm
\centerline{\it Lauritsen Laboratory of High Energy Physics}
\centerline{\it California Institute of Technology}
\centerline{\it Pasadena, CA. 91125}
\vskip.2cm
\author{Frank Wilczek\foot{Research supported in part by DOE
grant
DE-FG02-90ER40542}}
\vskip.2cm
\centerline{{\it School of Natural Sciences}}
\centerline{{\it Institute for Advanced Study}}
\centerline{{\it Olden Lane}}
\centerline{{\it Princeton, N.J. 08540}}
\endpage

\abstract{It is shown that
the breakdown of a {\it global} symmetry group
to a discrete
subgroup can lead to analogues of the Aharonov-Bohm effect.
At sufficiently low momentum, the
cross-section for
scattering of a particle with nontrivial $\Z_2$ charge
off a global vortex is almost
equal to (but definitely different from) maximal Aharonov-
Bohm
scattering; the effect goes away at large momentum. The
scattering of a spin-1/2 particle off a magnetic vortex
provides
an amusing experimentally realizable example.}

\endpage

The Aharonov-Bohm effect is generally thought to be
inextricably
connected to gauge symmetry, and to quantum
mechanics.  However, upon reflection there
are some funny aspects to these connections.
When the flux $\Phi$ is
expressed in terms of the fundamental unit $h/e$, so
$\Phi \equiv \tilde{\Phi} h/e$, and the scattered charge is
measured
in terms of the fundamental unit $e$, so
$q\equiv\tilde{q}  e$, then the
Aharonov-Bohm phase factor  $\exp (iq\Phi /\hbar )
=\exp(2\pi i\tilde{q}\tilde{\Phi})$
is independent of $e$ and $\hbar$.
This observation suggests that the Aharonov-Bohm effect
might
survive as $e$ and $\hbar$ approach zero, if the limit is
defined
in a suitable way.

We make these remarks not so much to outrage conventional
wisdom
concerning the Aharonov-Bohm effect, as to motivate
the possibility
of generalizing it.  Can something like it occur for
vortices of broken global symmetry, and in essentially
classical
contexts?   We shall argue here that indeed it can, and that
these generalizations have many potentially
interesting incarnations.

\chapter{Frame Dragging by Broken Symmetry}

To be definite let us consider
a model with global $U(1)$ symmetry broken down to
$Z_2$ by condensation of a scalar field $\lambda$.  Let
$\eta$ be a complex scalar field carrying half the $U(1)$
charge
of $\lambda$.  (For simplicity we shall assume that we are
dealing
with a relativistic theory; it will be clear that
the main points do not depend on this.)
Then generically one expects there to be
a coupling of the type
$$
\Delta {\cal L} ~=~ g \lambda \eta^2 + {\rm h.c.}
\eqn\framea
$$
In the homogeneous ground state where $<\lambda > = {v}$
this term generates a mass splitting between the
real and imaginary components of  $\eta \equiv (\rho_1 +
i\rho_2)/
\sqrt{2}~$:
$$
\Delta {\cal L} ~\rightarrow ~ {1\over 2}\Gamma (\rho_1^2 -
\rho_2^2)
\eqn\frameb
$$
where $\Gamma \equiv 2gv$.

Now in a vortex configuration for $\lambda$,
where $<\lambda (r, \phi )> ~\rightarrow ~ v e^{i\phi}$
outside a core region, it will still be possible to regard
the interaction \framea\ as generating a mass splitting
between
two real fields.  However as $\phi$ varies the orientation
of these
fields in internal space is dragged along---in fact,
rotated
by $\phi /2$.  In analyzing the dynamical effect of this
frame dragging, it is convenient to work with fields which
have a definite mass.  Thus let us introduce the local mass
eigenstates
$$
\tilde{\rho }=\left( \matrix { \rho_1 \cr \rho_2  } \right)
={1\over \sqrt{2}}\left( \matrix { e^{i\phi/2} & e^{-
i\phi/2} \cr
-ie^{i\phi/2} & ie^{-i\phi/2} \cr } \right) \left(
\matrix { \eta \cr \eta^* \cr } \right).
\eqn\transf
$$
(Strictly speaking this transformation will have to be
regulated near the origin.)
Now one can analyze the wave equation, to compute possible
scattering of excitations in the $\rho_i$ fields from the
vortex.  Because of the transformation \transf\ this wave
equation will have two unusual features:

\pointbegin
Each of the fields $\rho_i$ obeys the boundary condition
$$
\rho_i (\phi + 2\pi ) ~=~ - \rho_i (\phi ) ~.
\eqn\framed
$$
This means they must
be defined with a cut, or alternatively that the allowed
spectrum
of partial waves includes only {\it half-odd integers}.

\point
The gradient term $|\partial \eta |^2$ becomes modified, in
its
azimuthal component, to read
$$
{1\over 2 r^2} \bigl( (\partial_\phi
+ {i\over 2} \sigma_2 )\tilde{\rho } \bigr)^2~,
\eqn\framee
$$
where $\sigma_2$ is the Pauli matrix.
\par

What is the effect of these modifications?
We claim that at small
momenta $(k^2 \lsim \Gamma)$ the
second modification reduces to an additional potential
$$
V_{A^2} ~=~{1\over 4 r^2}
\eqn\framef
$$
Indeed
$\rho_1$ and $\rho_2$ have different effective
masses, and one should expect that perturbations connecting
them
are suppressed at small momenta.  Thus we can
neglect the terms linear in $\sigma_2$ at small
momenta.  (This is not quite obvious,
because the
$1/r^2$ interaction is potentially singular. However here
the
fact that the allowed partial waves are half-integral saves
the
day, because it means that there is always a centrifugal
barrier
shielding the origin.)
Thus the only significant effect of the
interaction with the vortex is to modify the boundary
conditions,
as in \framed\ , and to add an additional potential \framef.
We shall compute the resulting cross-section,
and justify our neglect of the off-diagonal terms, in the
following
section.  If we neglected the additional potential \framef\
then we
would have exactly the set-up which leads to maximal
Aharonov-Bohm
scattering.  The additional term introduces a calculable
modification,
which is relatively small for high partial waves (or small
angles).

On the other hand, clearly as $\Gamma \rightarrow 0$ the
effect
of the vortex must go away, apart from a possible
contribution from
ordinary scattering off the core (in the lowest partial
wave).  Thus
at large momenta $k^2 \gg \Gamma$ the induced
gauge field must essentially cancel the effect of the
modified boundary conditions.
Notice that the induced ``gauge field'' appearing in the
gradient energy, far from being responsible for
the Aharonov-Bohm-like scattering,
plays a crucial role in cancelling it off!

If the $U(1)$ broken symmetry {\it were} a
gauge symmetry, then the
``gauge field'' induced by the transformation \transf\ would
be
exactly cancelled by the true gauge field present in the
gauge covariant derivative of $\eta$ in the vortex
background.  Then we would have the classic Aharonov-Bohm
scattering
induced by the change in boundary conditions, at all
momenta.
Related to this, in a broken gauge theory the scattering
described
here, which (since it arises from the coupling to the scalar
Higgs
field) might appear to be additional to the
classic Aharonov-Bohm scattering,
in a sense reduces to an alternative representation of it.


Thus far we have considered the case of $Z_2$ charges.  For
higher global charges, a more complex situation emerges.
Consider for concreteness a $Z_3$ charge.  The interaction
corresponding to \framea\ is
$$
\Delta {\cal L} ~=~ g \lambda \eta^3 +{\rm h. c.}~.
\eqn\frameg
$$
The equation of motion for $\eta$ receives a contribution
of order $\eta^2$ from \frameg.  Thus, for small amplitudes
its effect is negligible.  In particular, there is no
scattering
from the $\lambda$ vortex, even for small momenta, in the
low
amplitude limit.  On the other hand for finite amplitude
waves
an analysis similar to the one given above applies.  For
small
momenta (where ``small'' now depends on the amplitude of the
wave)
it will be
appropriate to diagonalize \frameg, and one
will find the appropriate Aharonov-Bohm-like cross-section.
It is
of special interest to introduce quantum considerations at
this point.
If $\eta$ represents a quantum field, then the additional
interaction
\frameg\  represents a {\it vertex} where a single $\eta$
quantum
breaks up into two in the presence of a vortex
(or two fuse into one, or three appear from or annihilate
into
the vacuum).  Thus, for single quanta,
it appears quite different from an Aharonov-Bohm scattering
effect.
For coherent states of the $\eta$ field which can be treated
as
approximate solutions of the classical field equations,
however, the
preceding analysis applies.  Evidently the breakup and
fusion of individual
$\eta$ quanta induced by the vertex, induces space-time
deflection
of
their coherent superposition.


One might be concerned that, since the Nambu-Goldstone
excitations
associated with the broken symmetry field $\lambda$ are
massless
and therefore may be radiated with arbitrarily little
energy, the
effect discussed here could be washed out by Nambu-Goldstone
boson
emission.  However since the Nambu-Goldstone field is
derivatively
coupled, it is clear that on general grounds its emission is
an
order $(k/F)^2$ correction to the elastic process for small
momenta,
where $F$ is the scale of symmetry breaking. Therefore it
can be
made arbitrarily small in regimes of interest, and clearly
cannot
wash out the generic effect discussed here.


An essentially geometrical cross-section associated with a
global
symmetry poses a potential paradox; it is noteworthy how
this
paradox is resolved.
While
gauge charges have a universal coupling strength,
global charges do not, and so
it is difficult at
first hearing to understand how an essentially geometrical,
parameter-independent form of the cross-section could emerge
for
them.  Put another way, global charges have the character of
forbidding couplings, not mandating them---so how could
they
have a positive effect?
What we have seen is that there
is a geometrical cross-section
determined by the global charge, but the {\it domain of
validity}
of the cross-section, \ie\ the range in momenta for which it
is
valid, is a non-universal parameter that depends on the
strength
of the allowed coupling that fixes the charge
assignment.  As the strength of this
coupling goes to zero (removing, in principle, our ability
to define
the charge) the form of the cross-section remains unchanged
where it
is valid, but its range of validity shrinks to zero.

\chapter{Calculation}

Now we shall treat the prototype problem discussed above
more
quantitatively. For simplicity we will treat the
non-relativistic
case where the momenta $k$ of the incident particles
is much less than both of the perturbed masses
$\mu^2_{(1,2)}=
(m^2 \pm \Gamma)^{1/2}$.
We will also consider the quantum mechanical scattering
problem,
although similar considerations would apply to the classical
scattering of waves.

The substitution of eq.\transf\ into the equation for the
$(\eta,\eta^*)$
modes results in a non-relativistic Schr\"odinger equation
of the
form
$$
i\p_t \left( \matrix { \rhp  \cr \rhm \cr } \right)
=\left( \matrix { -{1\over 2\mup}(\del^2 - 1/4r^2) &
- {\p_{\phi}  \over 2\mup r^2} \cr
{\p_\phi \over 2\mum r^2} &
-{1\over 2\mum}(\del^2 - 1/4r^2) \cr } \right)
\left( \matrix { \rhp \cr \rhm \cr } \right),
\eqn\schro
$$
together with the boundary condition eq.~\framed.
The off-diagonal entries (terms
linear in the induced effective gauge field) connect states
of
different effective mass. Therefore, at low incident momenta
it is reasonable to expect that their effect will be small.
Our strategy will be to first solve the scattering problem
ignoring the off-diagonal terms, and then take them into
account
perturbatively. Of course, if we send in a pure $\rhm$ state
then there can be no real $\rhp$ production for incident
momenta below the threshold enforced by energy conservation.
However, even below this threshold, the off-diagonal
terms can affect the elastic scattering of the $\rhm$ modes
at second order in perturbation theory. We will argue below
that this effect is indeed small for incident energies much
less than the mass splitting.

The solution of the diagonal scattering problem proceeds
by performing a mode expansion (in two spatial
dimensions---or in three at normal incidence)
$$
\rhm (t,r,\phi) = \sum_{n\in\Z} e^{-i(\om+\mum) t}
e^{i(n+1/2)\phi} P_n^{(2)}(r),
\eqn\modeex
$$
(similar for $\rhp$). Note that the partial wave expansion
is shifted by one-half due to the boundary conditions.
Defining $x=\km r$ where $\om = (\km)^2/2\mum$ we find that
the radial eigenfunctions $P_n(x)$ satisfy a Bessel's
equation
of order $\nu_n^2 = (n+1/2)^2 + 1/4$.
The shift of $1/4$ from the usual order,
$\nu_n^2=(n+1/2)^2$, expected
with a mode expansion of the form eq.\modeex, is due to the
on-diagonal induced potential eq.\framef.

Thus the radial eigenfunctions that we will use for the
solution
of the scattering problem will be selected from the Bessel
and
Neumann functions $J_{\pm\nu_n}(\km r)$ and
$N_{\pm\nu_n}(\km r)$.
To select the appropriate set of solutions we must demand
square integrability of the solution near the position
of the vortex (taken to be the origin) and self-adjointness
of the Hamiltonian. This results in the selection
of only positive order Bessel functions in all angular
momentum modes {\it except} for $n=-1,0$.
As discussed (in the context of gauge strings) in the
appendix
of [\amrw ],
in both of these modes we have a one parameter family of
allowed boundary conditions corresponding to a mixture of
$J_{\nu_n}$
and $J_{-\nu_n}$. This apparent ambiguity results from the
unphysical
limit of zero core radius $R\to 0$.
The correct choice of boundary condition
is discovered by first performing a calculation at finite
$R$, and then
taking the limit $R\to 0$ [\amrw ].
In our case the correct choice is that we should
use only positive order Bessel functions in {\it all} modes.

Now we are
ready to construct the scattering solution and calculate
the elastic differential cross-section for incident
$\rhm$-modes.
This is most easily done if we reexpress the selected Bessel
functions in terms of outgoing ($H^{(1)}_{\nu_n}$) and
incoming
($H^{(2)}_{\nu_n}$) Hankel functions.
If we take the incident wave to be a plane wave $\exp(-i\km
x)$,
then we must construct out of the
Hankel functions a solution of the form
$$
\rhm^{\rm sol} = {1\over 2} \sum_{n\in \Z} e^{i(n+1/2)\phi}
e^{-i\pi\al_n/2}\left( H_{\nu_n}^{(2)}(k_2 r) +
H_{\nu_n}^{(1)}(k_2 r)
\right) ,
\eqn\formsol
$$
(note $J_n=(H_n^{(1)} + H_n^{(2)})/2$),
where $\nu_n$ is given by the positive square root.
The $\al_n$ are determined
by demanding that eq.~\formsol\ match onto
the incoming plane wave plus an outgoing scattered wave at
infinity;  we require
$$
\rhm^{\rm sol}\sim e^{i\phi/2}\left(
e^{-ik_2 x}+f(\phi){e^{ik_2 r}\over r^{1/2}}\right)~,
\eqn\ampdef
$$
where $f(\phi)$ is the scattering amplitude.
The phase $e^{i\phi/2}$ in front is necessary, because
of the double--valuedness of our solution, but it is
harmless---if we construct narrow wave packets that travel
in along the positive $x$-axis, then this phase is trivial.
Using the usual expansion of the plane wave
in terms of integer order Bessel functions
$$
\exp(-ikr\cos\phi)=\sum_{n\in\Z} e^{-i\pi |n|/2} e^{in\phi}
J_{|n|}(kr),
\eqn\planeex
$$
and the asymptotic behavior $H_\nu^{(1,2)}(x)\sim
(2/\pi x)^{1/2}\exp[\pm i(x-\nu\pi/2 -\pi/4)]$
of the Hankel functions,
the constraint of matching onto the plane wave determines
$\al_n=\nu_n$ for all $n$.

We can now calculate the phase shifts $\de_n(k_2)$ defined by
the asymptotic relation
$$
\rhm^{\rm sol} \sim {1\over 2} \sum_{n\in \Z}
e^{i(n+1/2)\phi}
e^{-i\pi|n|/2}\left( H_{|n|}^{(2)}(k_2r) + e^{i\de_n(k_2)}
H_{|n|}^{(1)}(k_2 r) \right).
\eqn\phasedef
$$
A simple calculation involving the asymptotic behavior of
the
Hankel functions leads to the result
$\de_n =  \pi(|n|-\nu_n)$.
{}From these phase shifts we can calculate the scattering
amplitude $f(\phi)$. The result is
$$
f(\phi)={e^{-i\phi/2}\over(2\pi i k_2)^{1/2}}
\left({1\over \cos(\phi/2)}
+2\sum_{n=0}^{\infty}(-1)^n(e^{i\Delta_n}-1)
\cos\left[\left(n+{1\over 2}\right)\phi\right]\right)~,
\eqn\ampexp
$$
where
$$
\Delta_n=\pi\left(n+{1\over2}-
\sqrt{\left(n+{1\over 2}\right)^2 +{1\over 4}}\right)~.
\eqn\modphas
$$
The first term in eq.~\ampexp\ is the usual maximal
Aharonov-Bohm
amplitude.  The corrections are due to the diagonal $1/4r^2$
potential,
and are most significant in low partial waves.  Retaining
only the first
term in the sum, we obtain the differential scattering
cross-section
$$
{d\si\over d\th}= {1\over 2\pi \km}\left( {1\over
\sin^2(\th/2)}
-8\sin^2(\pi(1-\sqrt{2})/4)\cos\th ~+~\cdots \right),
\eqn\diffcs
$$
where we have transformed to the correct scattering angle
$\th=\pi-\phi$.
Equations \ampexp-\diffcs\ are the final results of our
calculation.   We see that the differential scattering
cross-section of the
mass eigenstates off a global vortex (in this $\Z_2$
example)
is almost, but not exactly, of maximal Aharonov-Bohm form.

A calculation in second-order perturbation
theory shows that the effect of the neglected off-diagonal
terms
on the elastic scattering of $\rhm$ is bounded by
a constant times $(\km)^4/(m^2 \Gamma^2)$, uniformly in all
partial waves  [\promise ], and so can be neglected at low
incoming momenta.

\chapter{Examples}

\pointbegin{\it Spin 1/2 scattering by a magnetic vortex}

Consider a material with a planar magnetization
$\overrightarrow{M}(x)$---namely,
a material described by the XY model.  This model, of
course, supports vortices, which indeed play
an important role in its
dynamics.  A spin-1/2 particle, which might be an electron
or
a neutron for example, couples to the magnetization with
an interaction
$$
\Delta {\cal H} ~=~
g \psi^\dagger \overrightarrow{\sigma} \psi \cdot
\overrightarrow{M}~,
\eqn\exama
$$
where $\psi$ is the spinor field representing the particle.
The scattering of the spin-1/2 particle from the
magnetic vortex is an
instance of the general analysis above, but let us state it
in fresh terms.
In the presence of a vortex, the frame of the spin is
dragged around.
Thus if the magnetization is given by the vortex form
$\overrightarrow{M}{^i} (r,\phi ) \rightarrow
M_0 (-\delta_{i1} \sin \phi + \delta_{i2} \cos \phi )$,
then to keep the effective mass term generated by the
interaction
\exama\ diagonal, we shall need to transform to the
frame-dragged
variable $\tilde{\psi} ~\equiv~ \exp (i\phi \sigma_3/2
)\psi$.
Now as a spinor is rotated through $2\pi$, its sign
changes.  Thus for consistency, at low momentum where
parallel transport of the spin is appropriate,
the boundary condition on the spinor wave function
requires it contain only half-odd-integer angular momenta.
As
there also occurs an induced diagonal potential \framef, the
spin will scatter off the vortex with the cross-section
\diffcs.

The accumulation of phase by a spin subject to a magnetic
field
whose direction varies as a function of an external
parameter is
the classic example of Berry's
phase [\berry ].  The effect discussed here
may be considered in this framework, with the angle around
the
vortex playing the role of external parameter.
Indeed this point of view is instructive on several counts.
The restriction to low momenta we found above may be
considered as
the adiabatic condition for applicability of
Berry's phase---at
low momenta, the relevant trajectories (in the sense of a
Feynman path integral) for looping around the vortex
are traversed slowly.  Also the special role of the vortex
topology
is clarified---in circling the core, we surround a
point where an irremovable degeneracy between the
masses of $\rho_1$ and $\rho_2$, which are the eigenvalues
of the
local static Hamiltonia, occurs.  Finally various
generalizations,
such as to magnetizations tipped out of the plane
by angle $\be$ and sweeping out
a cone, suggest themselves.  That particular generalization
will change the calculation and cross-section as follows.
Upon diagonalizing the
interaction \exama\
we find that the effective mass term takes the form
$\tilde{\psi}^\dagger (\sin \be ~\sigma_3 +
\cos \be ~\sigma_2)\tilde{\psi}$,
with eigenspinors $\tilde \psi_\pm =
e^{-i{\be\over 2} \sigma_1 } (1, \pm i)^T$.
Now between {\it these} eigenspinors the effective gauge
potential proportional to $\sigma_3$ in the modified
gradient
term
$$
|\partial_\phi \psi |^2 ~=~
|(\partial_\phi - i{\sigma_3 \over 2} )\tilde{\psi} |^2~,
\eqn\examb
$$
{\it does\/} have non-vanishing diagonal matrix elements,
which
must be included in the calculation.  As a result, the
quantities $\nu_n$
are modified to become
$$
\nu_n^2 ~=~ (n + {1\over 2}) (n + {1\over 2} \pm \sin \beta)
+ {1\over 4}~,
\eqn\examc
$$
where the $\pm$ refers to the different eigenspinors.
{}From these the
cross-section is readily computed, but the formula is not
particularly
transparent.  It is noteworthy, however, that the leading
correction
to the canonical Aharonov-Bohm result contains terms in
$\sin \phi$
as well as $\cos \phi$, giving explicit parity and
time-reversal
asymmetries.

\point{\it Polarized light}

The essential requirement for the analysis of the previous
section
to apply is that there should be two degrees of freedom with
different
dispersion relations that are rotated into one another by
the
variation of a material parameter, such as a magnetization,
and
that when the material parameter rotates through a closed
cycle
each degree of freedom returns to itself, with a change of
phase.
This general set-up can be realized in a variety of optical
contexts,
where the degrees of freedom are two polarizations of light
of a given
frequency.
Realizations of frame-dragging for polarized light
have already been used for
interference experiments
[\tomita,\berryopt]; we are merely adapting it to
a realization in scattering.  Of course there is nothing
special about the optical region of the electromagnetic
spectrum
in this regard, and an alternative macroscopic realization
could
be constructed for microwaves propagating through ferrites.

\point{\it Passport to exotica}

Quite a few remarkable phenomena involving among others
Alice strings [\schwarz], Cheshire charge
[\cheshire,\prekra],
and flux tube-flux tube scattering [\ww,\bucher]
have been studied
in the context of spontaneously broken non-abelian gauge
symmetries.
Unfortunately, however, the list of spontaneously broken
non-abelian gauge symmetries
available for experimental manipulation is vanishingly
small.
The main point emphasized above,
that frame-dragging phenomena usually associated
with gauge theories also occur at low momenta for broken
global
symmetries, opens the strong
possibility that effects closely analogous to these can
be realized in suitable laboratory condensed matter systems.
Particularly interesting in this regard are helium 3
[\kman ] and liquid
crystals, which are known to have complicated order
parameter spaces
and to support non-abelian vortices [\mermin,\volovik].

Assuming the existence of a global
Aharonov-Bohm effect, it has been suggested
in the context of black hole physics that new forms of
``hair'' become measurable [\giddings ].
The effects we have described are important
for the
interaction of matter with global and axion strings, and may
affect their evolution in the early universe.

These matters are under active investigation [\promise].

\ack{We gratefully thank Mark Alford
for stimulating discussions.}

\par

\refout

\end